\documentclass[floatfix, twocolumn]{aastex62}
\usepackage{amsmath,amssymb,amsfonts,mathrsfs}
\usepackage{graphicx}
\usepackage{natbib}
\usepackage{color,soul}
\usepackage{hyperref}
\usepackage{comment}
\usepackage{afterpage}

\renewcommand{\sun}{$_\odot$}

\usepackage[english]{babel}
\usepackage[utf8x]{inputenc}
\usepackage[T1]{fontenc}
\usepackage{graphicx}
\usepackage[colorinlistoftodos]{todonotes}
\usepackage{tikz}
\usetikzlibrary{shapes.geometric, arrows}
\tikzstyle{startstop} = [rectangle, rounded corners, minimum width=3cm, minimum height=1cm,text centered, draw=black, fill=red!30]
\tikzstyle{io} = [trapezium, trapezium left angle=70, trapezium right angle=110, minimum width=3cm, minimum height=1cm, text centered, draw=black, fill=blue!30]
\tikzstyle{process} = [rectangle, minimum width=3cm, minimum height=1cm, text centered, draw=black, fill=orange!30]
\tikzstyle{decision} = [diamond, minimum width=3cm, minimum height=1cm, text centered, draw=black, fill=green!30]
\tikzstyle{arrow} = [thick,->,>=stealth]

\usepackage{rotating}
\usepackage{multirow}
\usepackage{hhline}

\newcommand{\grm}{\textit{galaxy-ranked}}
\newcommand{\Grm}{\textit{Galaxy-ranked}}

\newcommand{\iso}{\textit{PISO}}
\newcommand{\Msun}{$M_{\odot}$}

\graphicspath{{./}{figures/}}

\begin{document}

\title{Galaxy-Targeting Approach Optimized for Finding the Radio Afterglows of Gravitational Wave Sources}

\email{javed@iucaa.in}

\author{Javed Rana}
\affil{Inter University Centre for Astronomy and Astrophysics (IUCAA), S. P. Pune University Campus, Pune 411007, Maharashtra, India}

\author{Kunal. P. Mooley}
\altaffiliation{Jansky Fellow (NRAO, Caltech).}
\affil{NRAO, P.O. Box O, Socorro, NM 87801, USA}
\affil{Caltech, 1200 E. California Blvd. MC 249-17, Pasadena, CA 91125, USA}

\begin{abstract}
Kilonovae and radio afterglows of neutron star merger events have been identified as the two most promising counterparts, of these gravitational wave sources, that can provide arcsecond localization.
While several new and existing optical search facilities have been dedicated to finding kilonovae, factors such as dust obscuration and the daytime sky may thwart these searches in a significant fraction of gavitational wave events.
Radio-only searches, being almost immune to these factors, are equally capable of finding the counterparts and in fact offer a complementary discovery approach, despite the modest fields of view for many of the present-day radio interferometers. 
Such interferometers will be able to carry out competitive searches for the electromagnetic counterparts through the galaxy targeting approach.
Adapting and improving on an existing algorithm by \citeauthor{rana2017}, we present here a method that optimizes the placement of radio antenna pointings, integration time, and antenna slew. 
We simulate 3D gravitational wave localizations to find the efficacy of our algorithm; with substantial improvements in slew overhead and containment probability coverage, our algorithm performs significantly better than simple galaxy-rank-ordered observations.
We propose that telescopes such as the Very Large Array, MeerKAT, Australia Telescope Compact Array and the Gaint Meterwave Radio Telescope, having fields of view $\lesssim$1 deg$^2$ and searching for the counterparts of nearby GW events over tens of square degrees or larger, will especially benefit from this optimized galaxy-targeting approach for electromagnetic counterpart searches. 

\end{abstract}


\section{Introduction}\label{sec:intro}
The era of gravitational-wave (GW) astronomy has begun with the announcement of ten double black hole (BH-BH) mergers 
and the double neutron star (NS-NS) merger GW170817 \citep{GWTC1}.
Over the next few years ground-based GW detectors, such as the Advanced Laser Interferometer Gravitational-Wave Observatory \citep[aLIGO;][]{aasi2015}, Advanced Virgo \citep[AdV;][]{acernese2015} and KAGRA \citep{somiya2012}, will undergo sensitivity upgrades and thereby increase the number of compact binary coalescenses manyfold \citep{LRR_Prospects}. 
Detection of the first NS-BH merger is also imminent.
Once the GW detectors reach their design sensitivity, sometime between 2020--2022, they will have the NS-NS detection range\footnote{This is the average distance over the entire sky and orbital inclination for an SNR of 8. The horizon for detections (i.e. for optimally located and oriented sources) is a factor of $\sim$2.25 larger. During run O3, which lasts from 2019--2020, the NS-NS range will be about 120 Mpc for the two aLIGO detectors and 70 Mpc for the Virgo detector.} of 120--190 Mpc and BH-BH detection range of 1100--1600 Mpc \citep{LRR_Prospects}.
The NS-BH detection range is expected to be approximately twice that of NS-NS mergers.
At design sensitivities, the two aLIGO detectors together with AdV and KAGRA will be able to localize NS-NS mergers to $\sim$150 deg$^2$ \citep[median 90\% credible region;][]{LRR_Prospects}.
The addition of the LIGO-India detector, sometime after 2024, will reduce the median 90\% credible region for NS-NS mergers to $\sim$10 deg$^2$ \citep[70\% sources will be localized to within 20 deg$^2$;][]{LRR_Prospects}.

Identification of electromagnetic (EM) counterparts is essential for unlocking the full potential of gravitational wave discoveries \citep[e.g.][]{metzgerberger2012,nissanke2013}, although the large localization regions present a challenge for EM follow up.
Targeting of galaxies within the localization volume has been previously suggested \citep[e.g.][]{abadie2012,nissanke2013,gehrels2016} for significantly decreasing the sky area searched, and thereby making the most efficient use of observing time, even though the galaxy catalogs may be incomplete \citep[e.g.][]{dalya2018,kulkarni2018}. 
This is important when the telescope field of view is substantially smaller than the localization region, and always relevant for telescopes with $\sim$arcminute fields of view. 
The galaxy-targeting approach increases the chance of detecting the EM counterpart and reduces the number of false positives with respect to a blind search across the localization area \citep[e.g.][]{singer2016}.
This method has already been successfully demonstrated in the case of GW170817, which was localized to a 90\% credible area of 30 deg$^2$ \citep{abbott2017d}.

NS mergers (NS-NS and NS-BH) are considered to be the most promising GW sources having detectable EM counterparts \citep[e.g.][]{eichler1989,li_paczynski1998}.
There are two robust predictions for EM counterparts\footnote{Gamma-ray emission is also expected if the compact binary system is suitably oriented, but the gamma-ray localization for most mergers will be insufficient for determining a host galaxy.}: 1) thermal kilonova emission peaking at blue/optical or infrared wavelengths \citep[e.g.][]{metzger2010,barneskasen2013,kasliwal2017} on day--week timescales and 2) non-thermal afterglow emission, a power-law (generally optically thin) across the EM spectrum, which is suitable for searches at radio wavelengths \citep{nakarpiran2011,hotokezaka2016}.
The afterglow and kilonovae emission depend on different parameters: binary total mass, binary mass ratio, neutron star equation of state, geometry, and the energy and opening angle of the jet.
Radio and optical/IR searches for GW source counterparts have their own merits, and as such, offer complementary discovery approach. 
For some fraction of the NS mergers, the optical counterpart will be very difficult to detect with current survey instruments. 
This includes events localized in the daytime and bright-lunation sky and those lying in dust-obscured environments (about 30\% of all GW events), NS-NS mergers having large total mass, and a large fraction of the NS-BH mergers \citep[e.g.][]{kasliwal2016,kasen2017}.
While the kilonova signal is short-lived, radio afterglows evolve on longer timescales \citep[weeks to years;][]{nakarpiran2011,hotokezaka2015,hotokezaka2016,hallinan2017,mooley2018-strongjet}, allowing ample time for discovery and follow up.
There are also some challenges in the radio follow up of GW triggers, such as
the modicum of radio facilities available (at least compared to the optical ones), and 
the uncertainties in the density of the circum-merger environment that determines the intensity of radio emission.
Therefore, the radio may play a follow-up role on several occasions, as in the case of GW170817, but at other times the radio may be the only way to localize the EM counterpart. 
The latter cases motivate optimized schemes for radio-only searches across the GW (or GW + gamma-ray) localization regions. 

Previously-proposed telescope scheduling/observing schemes have focused on optical follow up \citep{rana2017, ghosh16, Coughlin:2018lta, salafia17} of GW sources or searches with wide-area radio detectors (ASKAP; Dobie et al. 2019, submitted). 
Here, we develop an optimal scheme for finding the radio afterglows of GW sources via the galaxy-targeting approach.
This technique will be important for radio interferometers, like the Karl G. Jansky Very Large Array (VLA), the Australian Telescope Compact Array (ATCA), the upgraded Giant Meterwave Radio Telescope, and MeerKAT, having fields of view $\lesssim$1 deg$^2$ and searching for the counterparts of nearby GW events over tens of square degrees or larger.
We adapt and improve on an existing algorithm \citep[note also the combined tiling and galaxy-targeted search strategy proposed by \cite{Rana2019a};][]{rana2017} which took into account the rising/setting time, airmass, and sun/moon constraints.
In this work, we incorporate slew and exposure time optimization, taking into account the shape of the primary beam of a typical radio interferometer such as the VLA. 

This paper is organized as follows.
In \S\ref{sec:algo} gives a mathematical and flowchart descriptions of our radio search algorithm.
In \S\ref{sec:sim} we describe the simulation of GW localizations which are then used to find the efficacy of the search algorithm in \S\ref{sec:application}.
\S\ref{sec:mitigation} explores a means to mitigate the effect of galaxy catalog incompleteness.
We present the summary and possible improvements to our algorithm, that can be made in the future, in \S\ref{sec:summary}.

We note beforehand that galaxy catalog completeness and galaxy weighting schemes are not analyzed in detail in this work.
However, we do provide some guidelines for curcumventing the catalog completeness issue, and allow the galaxy weighting (e.g. $X\times$SFR + $Y\times$Mass) to be adjusted by the users of our algorithm.

\section{Algorithm}\label{sec:algo}

In this section we discuss the algorithm to optimize the radio observations.
We call this algorithm as \textit{Pointing, Integration-time and Slew Optimization (\iso\footnote{We have made our \iso\ Python code (updated regularly) available via GitLab.})} method. Our starting point is the 3D probability distribution of the GW source location.
The probability distribution over the full sky is normalized as follows, $p(\hat{r})$ being the probability density at radial vector $\hat{r}$.
\begin{equation}
\int_{full sky} p(\hat{r}) d\hat{r} = 1
\end{equation}

Given a galaxy catalog, we measure the mass of each galaxy from its optical (e.g. B band) luminosity and distance. We use the masses of the galaxies to put different weight to the galaxies. However, the galaxy weight can depend on the other parameters of the galaxy. As an example, along with the mass of the galaxy, the star formation rate of the galaxy can be used to put the weight for that galaxy. The algorithm is independent of the way the galaxy weights are calculated. The galaxy weights from different methods can be adopted easily in our code.  

Consider the probability of the unit volume around the galaxy $g_i$ is $p(\hat{r}|g_i)$, then the galaxy-weighted weighted normalized probability becomes 
\begin{equation}
\int_{full sky} p(\hat{r}|g_i)\: m_i\: d\hat{r} = 1
\end{equation}
Where $m_i$ is the mass of the $ith$ galaxy.

The optimizing scheme contains two independent procedures: 1) optimizing the locations of the antenna pointings, i.e. where to point the telescopes given a particular galaxy containment probability distribution on the sky, and 2) optimizing the telescope slew between the pointings. We describe these procedures individually in the following subsections.

\subsection{Optimizing the locations of antenna pointings and integration time}\label{sec:algo:math}

The radio antenna primary beam has a sensitivity function\footnote{Generally a Gaussian or power-law approximation, which is fairly accurate down to a few percent of the sensitivity function, is used to describe the primary beam. For the purposes of this work, we define "galaxies contained within the beam" as "galaxies out to $\sim$10\% power point of the primary beam". Galaxies beyond this distance from the pointing center are not considered.} $b(\alpha)$, where $\alpha$ is the angular distance from the center of the beam. The galaxy-weighted map and the beam function are convolved to produce a convolved map of the full sky. In this convolution, the beam probability is calculated putting the beam center at every pixel of the GW localization map (see \S\ref{sec:sim} for further details). The probability of the beam is the sum of the probability of the galaxies within it.
Consider a beam, centered at $(\theta, \phi)$ on the sky, containing $n$ number of galaxies and the probability of the beam is $p_b$.

\begin{equation}
\int_{beam} p(\hat{r}|g_i)\: m_i \: b[(\theta_j, \phi_j), \alpha_i] \: d\hat{r} = \sum_{beam,i} p_b(\theta_j, \phi_j, r_i) 
\label{equ:beam_probability}
\end{equation}

\begin{equation}
\int_{full sky} p(\hat{r}|g_i)\: m_i\: b[(\theta_j, \phi_j), \alpha_i] \: d\hat{r} = \sum_{full sky} p_b(\theta_j, \phi_j, r_i)
\label{equ:beam_int}
\end{equation}

Where $(\theta_j, \phi_j)$ is the co-ordinate of the center of the $jth$ beam and $r_i$ is the distance of the $ith$ galaxy from the Earth. Eq.~\ref{equ:beam_int} defines a grid of pointings that we optimize below, such that the containment probability is maximized.

Our goal is to carry out a sensitive search over all the galaxies, achieving a limiting radio luminosity $\mathscr{L}$.
Now, $\mathscr{L} \propto$ ($rms$ noise)/(distance)$^2$  and $rms \propto 1/\sqrt{T}$, so the 
integration time for one beam, pointing at $(\theta, \phi)$ on the sky, depends on the farthest galaxy within the beam as,
$T(\theta, \phi) \propto d^{4}_{max}$ 
where $d_{max}$ is the distance (from Earth) of the farthest galaxy within the beam.
If the total number of pointings within the available observing time, $T_{too}$ (ToO=Target of Opportunity), is $N$ then $\sum_{j=1}^{N} T_j = \sum_{j=1}^{N} T(\theta_j, \phi_j) = T_{too}$. In that case, the total probability covered within the $T_{too}$ time is $P = \sum_{T_{too}} p_b(\theta_j, \phi_j) = \sum_{j=1}^{N} p_b(\theta_j, \phi_j)$.

To minimize the beam integration time and maximize the probability we put equal weights to the $T(\theta, \phi)$ and $p_b(\theta, \phi)$, i.e. we simultaneously optimize the containment probability and the integration time, giving equal importance to both parameters.
For each beam described by equation~\ref{equ:beam_int}, and located at the center of each pixel of the localization map, the center of the beam is shifted in such a way that it maximizes the containment probability within the beam (described by equation~\ref{equ:beam_probability}).

\begin{equation}
P_{max} = Max\Big(\sum_{T_{too}} p_{b}[\theta_j+\delta\theta, \phi_j+\delta\phi]\Big) \quad \forall \; \delta\theta < r_b \;\&\; \delta\phi < r_b
\label{equ:beam_shift}
\end{equation}
Where $r_b$ is the angular radius of the beam. $\delta\theta$ and $\delta\phi$ are the displacements of the center of the beam along $\theta$ and $\phi$ respectively.

\subsection{Slew optimization}\label{sec:algo:slew}
In this subsection we explain the slew optimization algorithm that minimizes the total amount of slew under the one epoch of follow-up. Optimizing slew is similar problem as the minimization of the total cost of travel in traveling salesman problem (TSP). One main difference is that the start and end points in the TSP are same, whereas in slew minimization case the start and end are two separate points. In most of the cases, the slew angle depends on the geometry and the size of localization. In practice, observer prefers to order the imaging of the galaxies according to their probabilities, but as their positions in the localization are random the total slew angle is not minimized at all. Here we propose a method to minimize the slew using the algorithm called ``Nearest Neighbor and Local Search'' (NNLS) . The ``Nearest Neighbor'' (\href{https://en.wikipedia.org/wiki/Nearest_neighbor_search}{NN}) search gives an approximate solution to the slew optimization problem. Sometime the algorithm gives a solution far from the optimal solution. So we add the local search with the NN algorithm. The NNLS algorithm looks for the nearest neighbor point and also the local points close to it. We found the NNLS algorithm is efficient in minimizing the slew because of the arc-shaped GW-localization on the sky. There are other algorithms to solve the TSP more accurately, which are computationally expensive\footnote{The exact solution by brute force method has a computational factor $(n-1)!$, where n is the number of galaxies. On the other hand, in NNLS the computational factor is $O(n^3)$. In this method, it follow the "Nearest Neighbor" (NN) method but also search for the other local points.} compared to NNLS. As the computation time is one of the main consideration for transient search, we prefer NNLS. We note that this approach is not exactly the {\emph optimal} solution, but it is close to optimal. 

\subsection{Flowchart}\label{sec:algo:flowchart}

Figure~\ref{fig:flowchart} presents the flowchart of the \iso\, method.
The description of the flowchart is given below:   

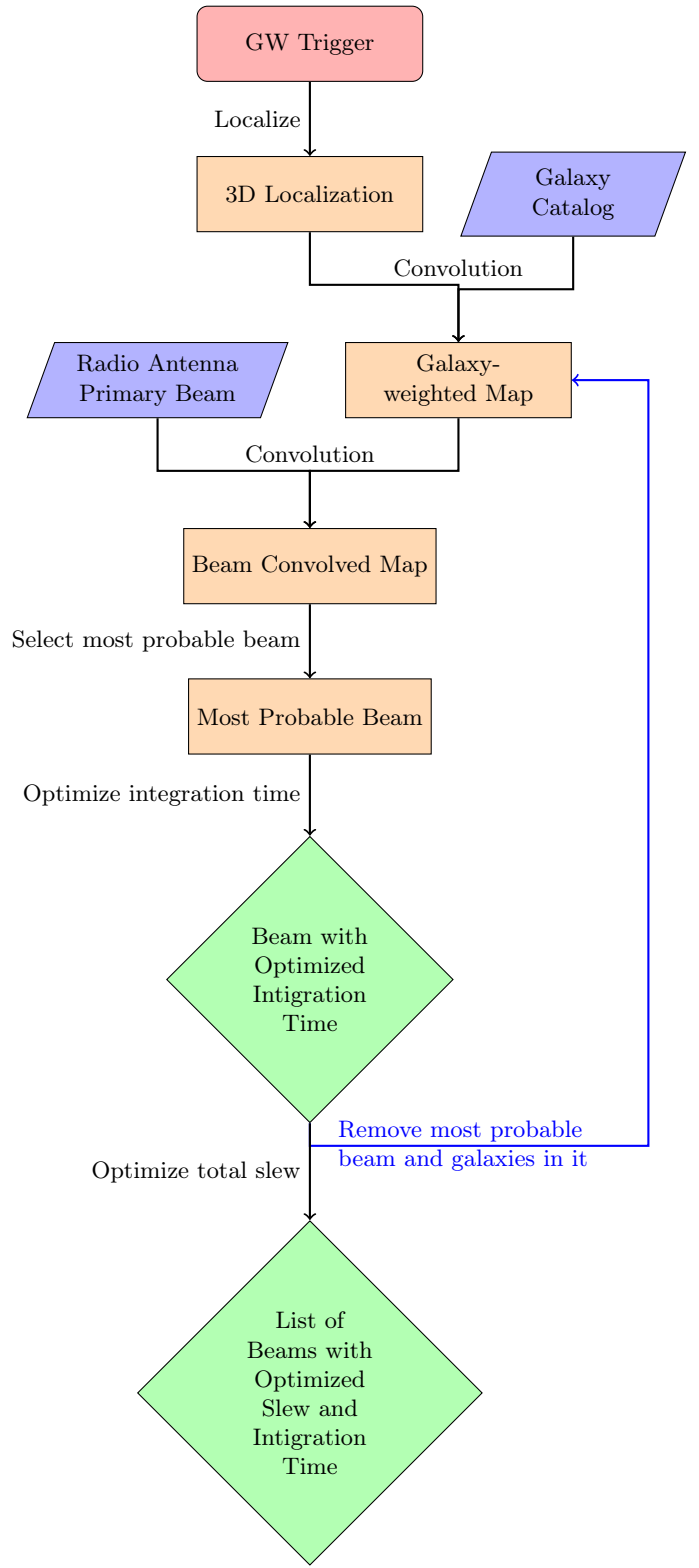
\begin{figure}[htp]
\begin{tikzpicture}[node distance=2cm]
\node (gwtigger) [startstop] {GW Trigger};
\node (3dlocalization) [process, below of=gwtigger, text width=2.5cm] {3D Localization};
\node (gc) [io, right of=3dlocalization, text width=1.5cm, xshift=1.5cm] {Galaxy Catalog};
\node (3dconvomap) [process, below right of=3dlocalization, node distance=3.5cm, text width=2cm, xshift=-.5cm] {Galaxy-weighted Map};
\node (beam) [io, left of=3dconvomap, text width=2.2cm, xshift=-2cm] {Radio Antenna Primary Beam};
\node (beamconvolved) [process, below left of=3dconvomap, node distance=3.5cm, xshift=.5cm] {Beam Convolved Map};
\node (mpbeams) [process, below of=beamconvolved] {Most Probable Beam};
\node (expoptim) [decision, below of=mpbeams, text width=2cm, node distance=3.5cm] {Beam with Optimized Intigration Time};
\node (slewoptim) [decision, below of=expoptim, text width=2cm, node distance=5.5cm] {List of Beams with Optimized Slew and Intigration Time};

\draw [thick, ->] (gwtigger) -- node[anchor=east] {Localize} (3dlocalization);
\draw [thick, ->] 
	(gc.south) - ++(0,-0.7cm) -| (3dconvomap);
\draw [thick, ->] 
	(3dlocalization.south) - ++(0,-0.7cm) -| node[above] {Convolution} (3dconvomap);

\draw [thick, ->] 
	(beam.south) -- ++(0,-0.7cm) -| (beamconvolved);
\draw [thick, ->] 
	(3dconvomap.south) -- ++(0,-0.7cm) -| node[above] {Convolution} (beamconvolved);
    
\draw [thick, blue, ->] 
	(expoptim.south) -- ++(0,-0.3) -- ++(4.5,0) node[above, text width=4cm, anchor=east, node distance=3.5cm] {Remove most probable beam and galaxies in it} |- (3dconvomap.east);

\draw [thick, ->] (beamconvolved) -- node[anchor=east] {Select most probable beam} (mpbeams);
\draw [thick, ->] (mpbeams) -- node[anchor=east] {Optimize integration time} (expoptim);
\draw [thick, ->] (expoptim) -- node[anchor=east] {Optimize total slew} (slewoptim);

\end{tikzpicture}
\caption{Flowchart of the \iso\ galaxy-targeting algorithm.
\label{fig:flowchart}}
\end{figure}

\begin{enumerate}
    \item \textit{GW-localization}: The 3d sky localization of a GW event is generated from the GW trigger parameters of the network of detectors. BAYESTAR \citep{Bayestar} generates the sky localization in HEALPix \citep{HEALPix-paper} format, where the sky is divided in pixels. All the pixels are equal in angular area. Every pixel has a probability distribution along the radial direction.
    
    \item \textit{Galaxy Catalog}: 
    The galaxy catalog, having right ascension (RA), declination (Dec) (and Mass, SFR etc. as needed for the galaxy weighting), chosen for convolving with the GW 3d-localization
    
    \item \textit{Galaxy-weighted map}: The special convolution of the 3d probability distribution in the GW-localization and the galaxy distribution within it gives the galaxy-weighted map. In the convolution process, we weight each galaxy by its mass (higher the mass higher the weight). We normalize the total weight of the convolved map to make it a probability distribution. The convolved map is a distribution of discreet points (the optical centers of the galaxies) in the 3-dimensional space. We assume that, if a pixel has no galaxy within it then the probability of finding the radio counterpart of the GW source in that pixel is zero (but see \S\ref{subsec:comparison} in the context of galaxy catalog incompleteness).
    
    \item \textit{Radio antenna beam}: For simplicity, the beam of the radio antennas is assumed to have azimuthal symmetry. The beam has maximum sensitivity at the center of the pointing. The sensitivity at any other point on the beam decreases as its angular separation from the center increase. 
    
    \item \textit{Beam convolved map}: To generate the beam convolved map we perform the special convolution of the radio antenna beam and the galaxy-weighted map. The probability of the convolved map is calculated by putting the beam-center at every pixel in the map. The probability of a beam is the sum of the probability of the galaxy-weighted points within the beam.
    
    \item \textit{Most probable beam}: 
    The integration time for the observation of a beam depends on the most distant galaxy within the beam. If there is a very distant galaxy, then the integration time to observe the beam might be very long and that can consume most of the ToO time. So, we put an upper cap in the integration time to avoid such cases. In our simulation, we put equal weight to the probability and the integration time to choose the most probable beam.
    
\begin{figure}[thbp]
  \includegraphics[trim=1cm 0 0cm 0cm, clip=false, width=0.5\textwidth]{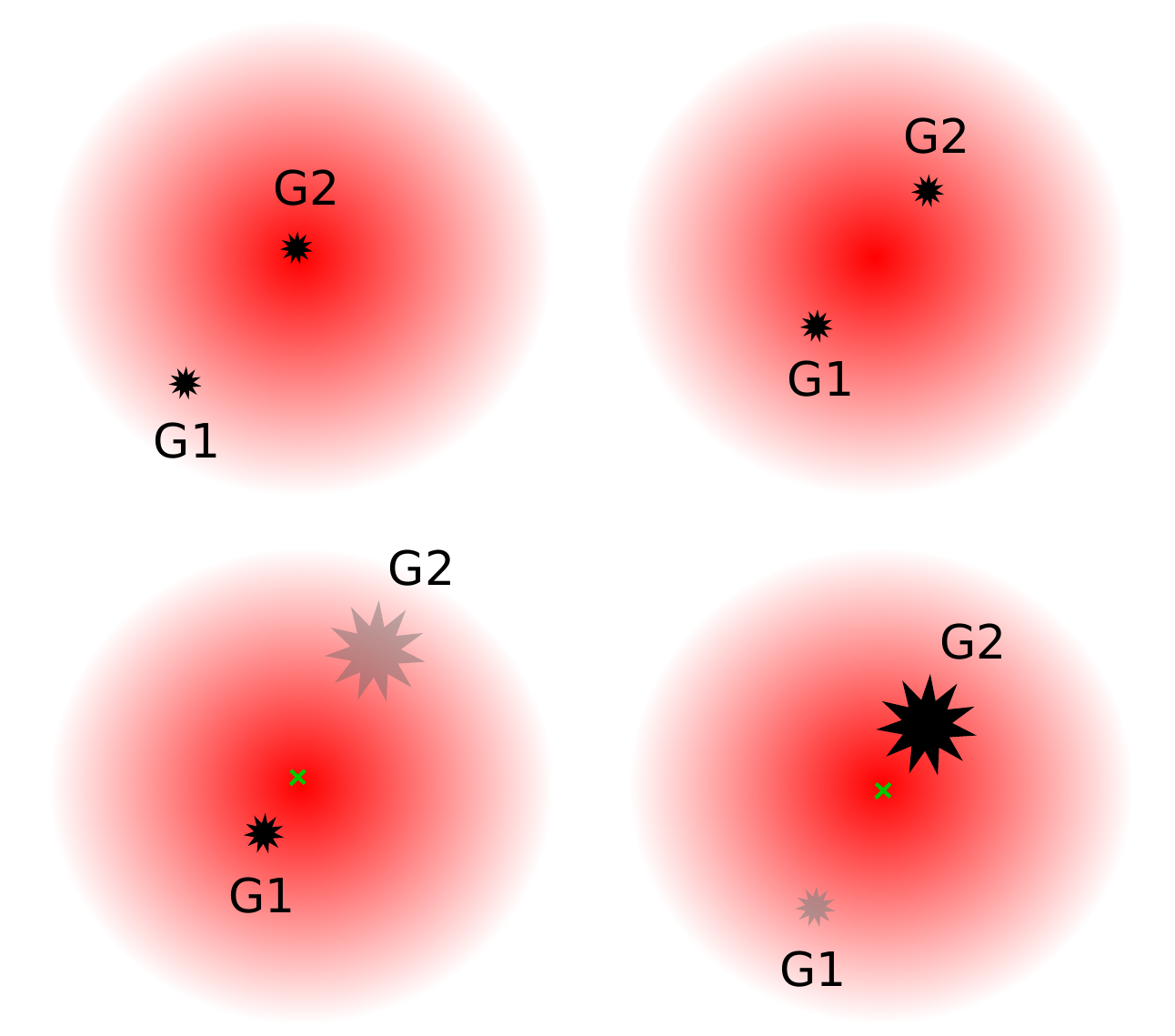}\hspace*{0.5cm}
  \caption{A simple example illustrating the optimal placement of the radio antenna primary beam. The red shaded region is the primary beam and the density of the color represent the sensitivity. The two black points are two galaxies (G1 \& G2) at the same distance from earth within the beam. The size of the dot is inversely proportional to the distance of the galaxy, bigger the size closer the galaxy. The color of the dots represents the containment probability of the galaxy, darker the color higher the probability. The green cross is the center of the beam. Top-Left: One galaxy (G2) is at the center of the beam, where the sensitivity is maximum. The other galaxy (G1) is at a point where the sensitivity of the beam is half of the maximum sensitivity. Top-Right: The beam center is placed at the middle of the angular separation of the two galaxies, such that the sensitivity of the radio antennas at the two galaxies are same. Bottom-Left: G1 galaxy is at farther distance with higher probability than G2 galaxy. In this case, the optimal beam pointing is such that the center of the beam is closer to the G1 galaxy. Bottom-Right: G1 galaxy is at farther distance with lower probability than G2 galaxy. In that situation, the optimal beam pointing is such that the center of the beam is closer to the G2 galaxy.
  \label{fig:radio-pointing}}
\end{figure}

    \item \textit{Optimize the integration time}: As the sensitivity of the antennas are not uniform within the beam, the small shift of the center of the beam might reduce the integration time to observe it. As a simple example, consider a beam with two galaxies within it, G1 \& G2, as shown in Figure~\ref{fig:radio-pointing}. The size of the dot is inversely proportional to the distance of the galaxy: bigger the dot closer the galaxy. The color shade of the dots represents the containment probability of the galaxy: darker the dot, higher the probability. 
    In Figure~\ref{fig:radio-pointing}, top-left: G2 galaxy is at the center of the beam and G1 galaxy is at the edge of the beam. In this case, the beam center is not optimally placed. On the other hand, we can sift the center of the beam (under the constraint $\delta\theta < r_b, \delta\phi < r_b$ as described by equation~\ref{equ:beam_shift}) to the middle of the two galaxies, such that the sensitivity of the antennas are the same at both the galaxy points. To detect a radio source in the G1 galaxy at the top-left panel of Figure~\ref{fig:radio-pointing}, beyond the luminosity threshold $\mathscr{L}$ and above the fixed SNR threshold (e.g. 4$\sigma$), 
     will take longer integration time than the radio source being in the G1 galaxy in the top-right panel of figure~\ref{fig:radio-pointing}. Bottom-Left: G1 galaxy is at farther distance with higher probability than G2 galaxy. In this case, the optimal beam pointing is such that the center of the beam is closer to the G1 galaxy. Bottom-Right: G1 galaxy is at farther distance with lower probability than G2 galaxy. In that situation, the optimal beam pointing is such that the center of the beam is closer to the G2 galaxy.
    
    \item \textit{Repeat step-4 to step-6}: Once we have the most probable beam after minimizing the integration time, we remove the most probable beam, and the galaxies within that beam from the galaxy-weighted map. This gives a new galaxy-weighted map with a total probability less than unity. We repeat the steps from 4 to 6 until the full ToO time is exhausted. At the end, we get certain number of pointongs to observe based on the total ToO time. 
    \item \textit{Slew optimization}: Once the integration time associated with all pointings is calculated, we use the NNLS method is used to minimize the total amount of slew of the radio antennas. The pointing that is expected to set (below horizon or the minimum possible elevation/altitude) earlier than the other pointings is taken as the first pointing observed. Thereafter, the order of the pointings is determined by the setting time of the other pointings and slew minimization. 
\end{enumerate}

Finally we get an order of observation with optimized integration time and minimized total slew.
We note that, in rare cases a most-probable beam may contain a very distant galaxy (compared to the others in the beam), and the integration time for that beam might increase substantially, thus consuming a large fraction of the ToO time. So, our algorithm offers a user-specified upper cap on the integration time to be placed, in order to avoid such situations.

\begin{figure*}[thbp]
  \centering
\includegraphics[trim=1cm 0 0cm 0cm,clip=false,width=2.2in]{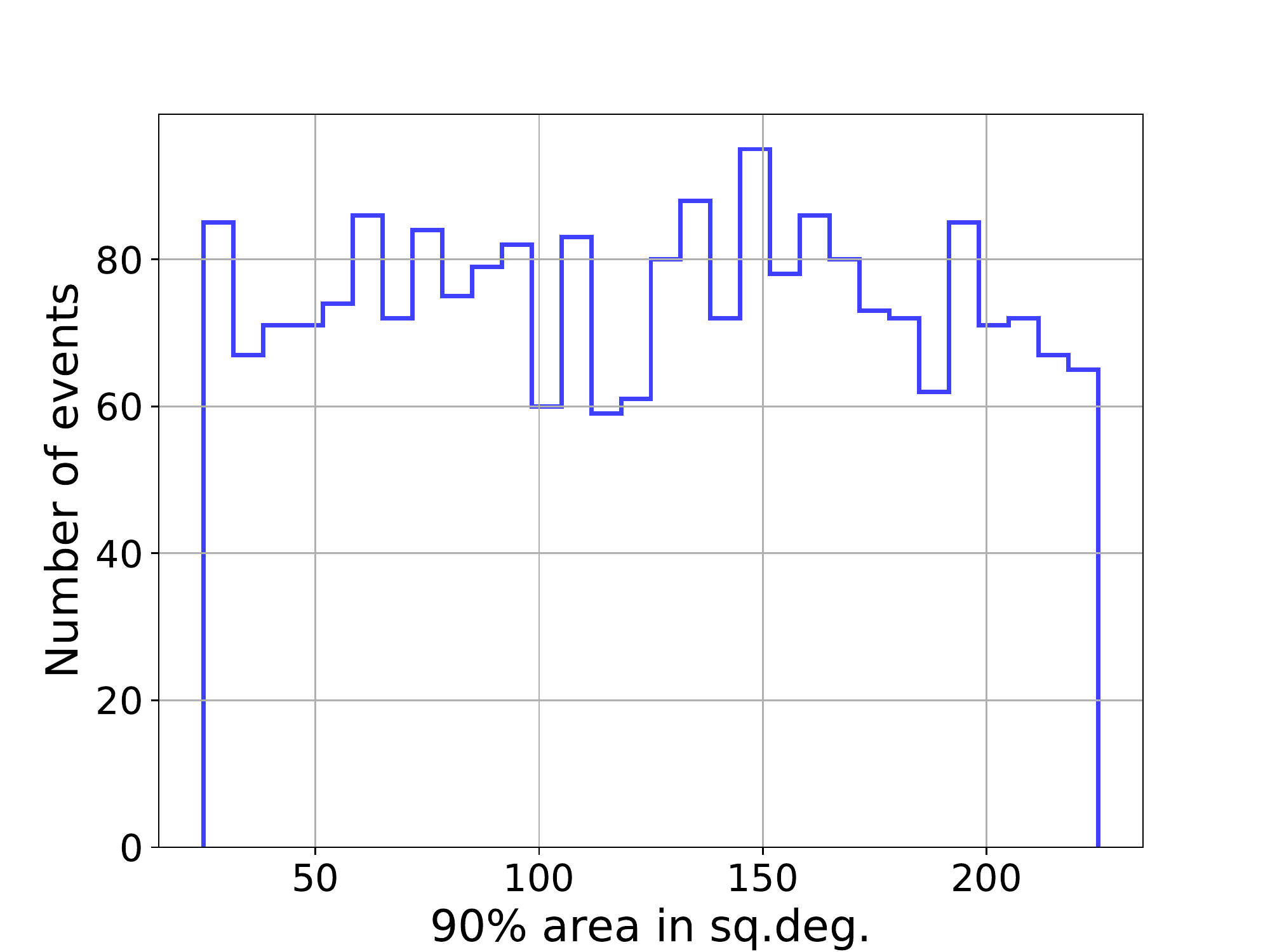}
\includegraphics[trim=1cm 0 0cm 0cm,clip=false,width=2.2in]{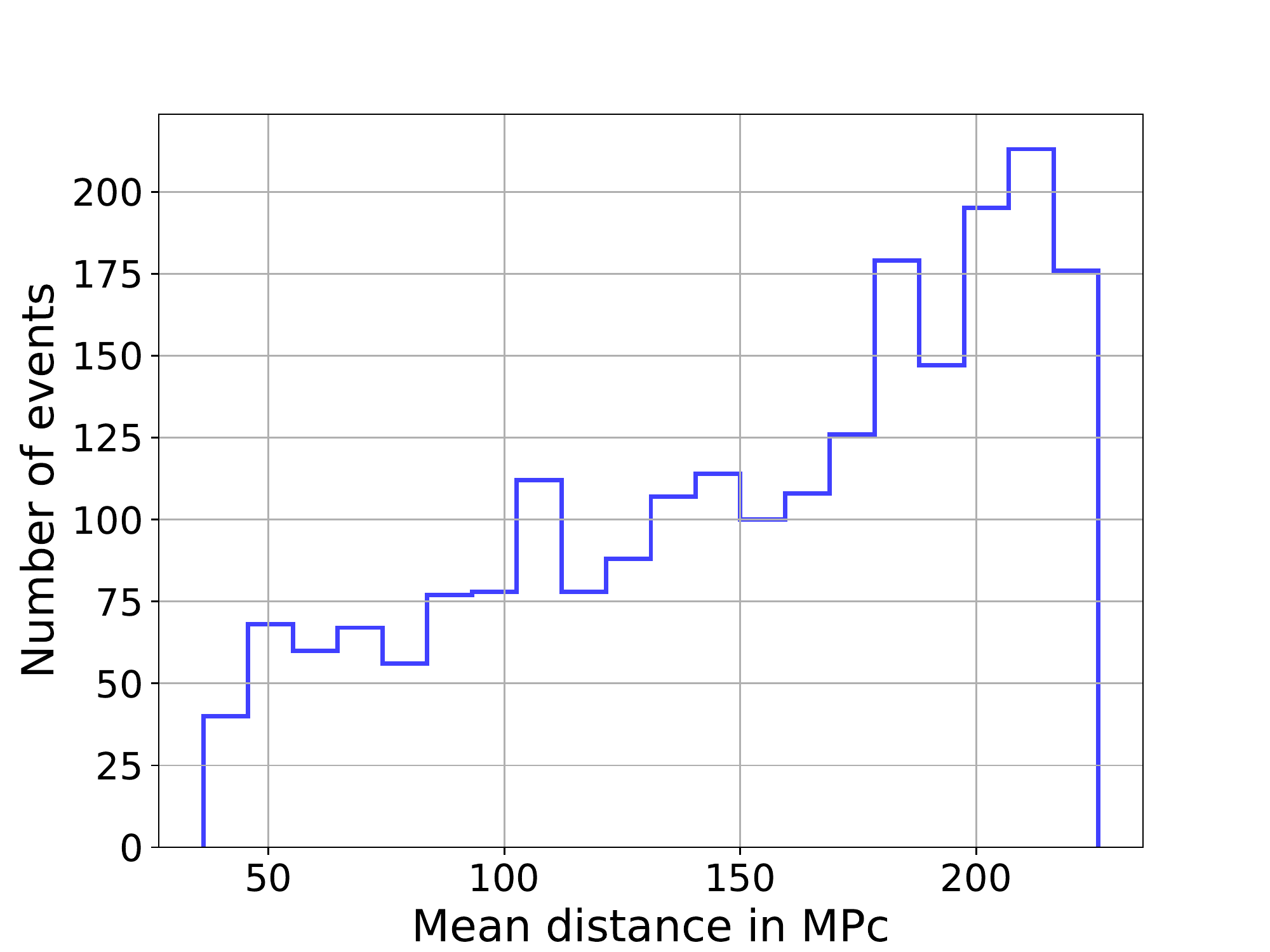}
\includegraphics[trim=1cm 0 0cm 0cm,clip=false,width=2.2in]{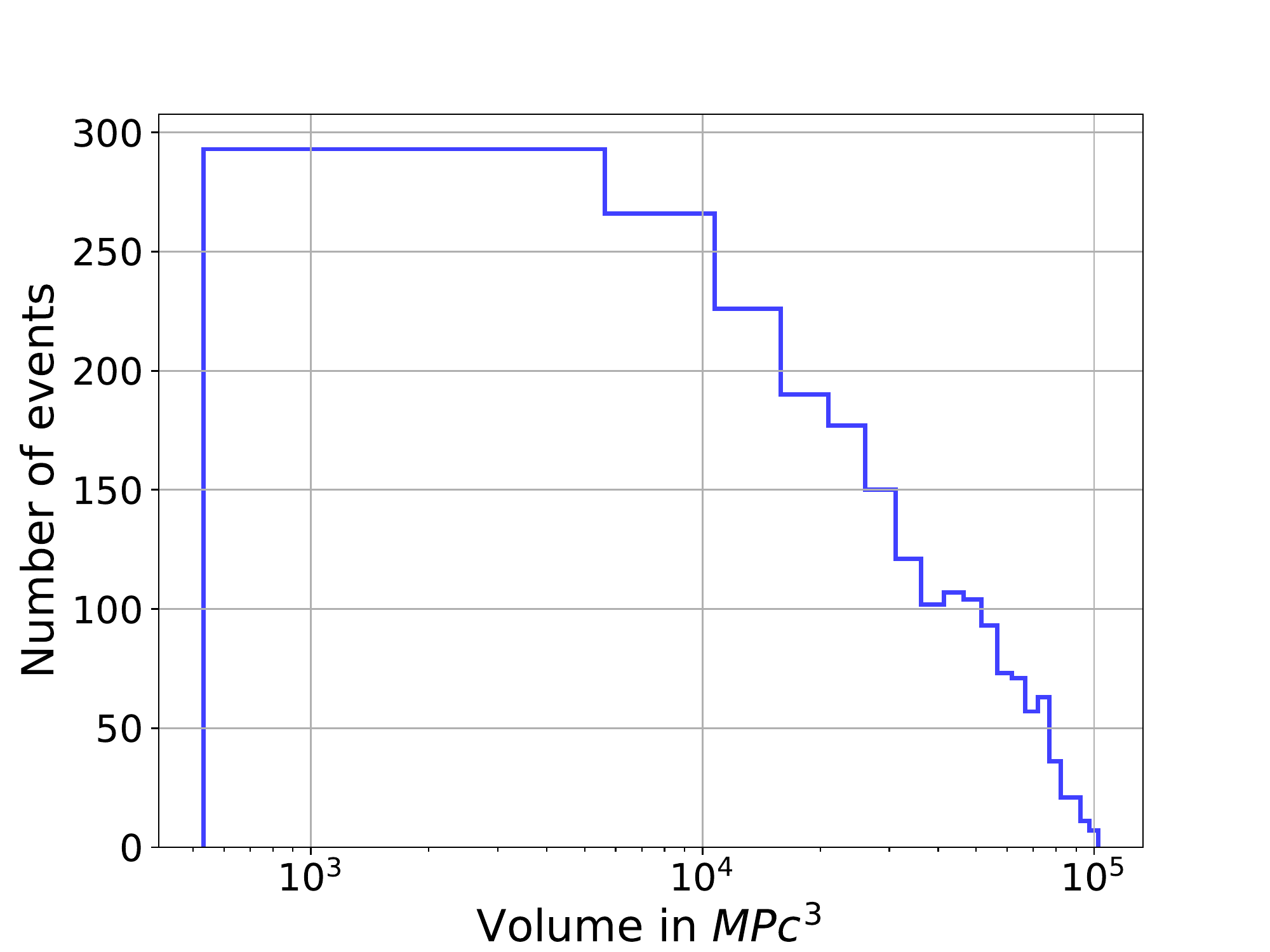}
\caption{Visual summary of the $\sim$2000 GW localizations (NS-NS merger events) simulated. See \S\ref{sec:sim} for further details on the simulation. Left: The distribution of area of the 3D-localizations. The X--axis is the area of the 90\% probability containment region and the Y--axis shows the number of events.
Middle: The distribution of distance of the events. The X--axis is the area of the 90\% probability containment region and the Y--axis shows the number of events.
Right: The distribution of volume of the 3D-localizations. The X--axis is the area of the 90\% probability containment region and the Y--axis shows the number of events.}
\label{fig:sim}
\end{figure*}

\section{Simulating Merger Events}\label{sec:sim}
In order to test (and find the efficacy of) our radio follow-up algorithm on GW events, we simulated $\sim$2000 GW localizations.
We used NS-NS mergers since these are the most promising GW events with EM electromagnetic counterparts,
In our simulation, we focus on three detector's network (HLV) in their design sensitivities \citep{AdvLIGO, AdvVIRGO}. We keep the component masses of all the binaries at 1.4 \Msun. We consider the binaries are non-spinning systems\footnote{Spin will not affect the localization volume substantially ($\sim$10\% or less). Spin templates are computationally expensive.}. Source positions are random, isotropic, and uniform in volume. The source distance has a cut off at 225 Mpc, based on the completeness of the galaxy catalog and the galaxy saturation in the beams. We used BAYESTAR \citep{Bayestar} and LALSuite \citep{LALSuite} to generate the volume localizations, i.e. 3D probability distributions \citep{singer2016}.
BAYESTAR saves the localization data in HEALPix\footnote{A HEALPix map of $N_{side}$, with $N_{pix}=12\times N_{side}^2$ is a pixelized representation the full sky. Every pixel in the map has 1) a containment probability and 2) probability distribution, $p(\hat{r})$, along the distance axis $\hat{r}$. The latter is usually considered to be a Gaussian distribution \citep[e.g.][]{singer2016}. The probability density is therefore a function of the 2D location on the sky and the distance, $p=p(\theta, \phi, r)$.}
The area, volume and distance distributions of the simulated GW  localizations are shown in Figure~\ref{fig:sim}.
The localizations have been generated randomly on the sky, and $\sim$1200 (64\%) of them have their 90\% containment regions above a declination of $-40$ deg (and are therefore observable with the VLA, for example). 

\begin{table*}[hbtp]
\caption{Search Methods\label{table:methods}}
\begin{center}
\begin{tabular}{|l|c|c|c|c|}\hline
\textit{Methods} & Integration time & Beam center & Slew & Setting \\\hline
\Grm\  & Not optimized & Not optimized & Not minimized & Not considered \\\hline
\iso\  & Optimized & Optimized & Minimized & Considered \\\hline
\end{tabular}
\tablecomments{Comparison between the two radio transient search methods.}
\end{center}
\end{table*}

\section{Algorithm Application to Simulated Merger Events}\label{sec:application}
In this section we demonstrate the efficacy of our algorithm to select and sequence pointings based on the galaxies within the localization to maximize the containment probability
(given the available observing time).

We consider an intuitive strategy for galaxy targeting: sequentially pointing at galaxies that are rank-ordered (according to their GW source containment probabilities, star formation rates, mass etc.). 
We call this the \grm\ method, and compare results from our algorithm (\iso) with those from this simplistic method.
In Table~\ref{table:methods} we compare the parameters optimized in each of these two methods.

\makeatletter\onecolumngrid@push\makeatother
\begin{figure*}[t]
  \centering
    \includegraphics[trim=0cm 0cm 0cm 1cm, clip=true, width=0.45\textwidth]{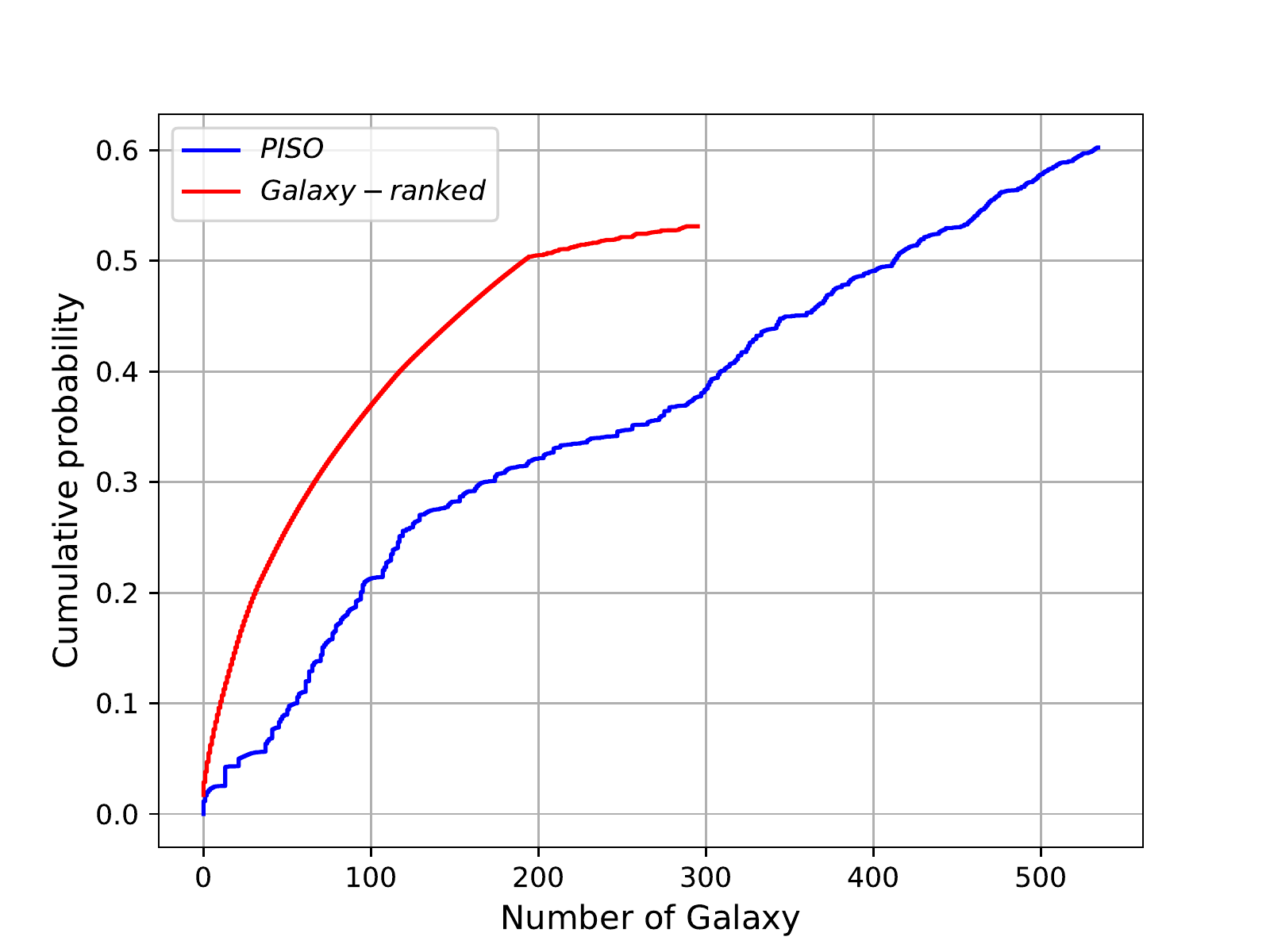}
    \includegraphics[trim=0cm 0cm 0cm 1cm, clip=true, width=0.45\textwidth]{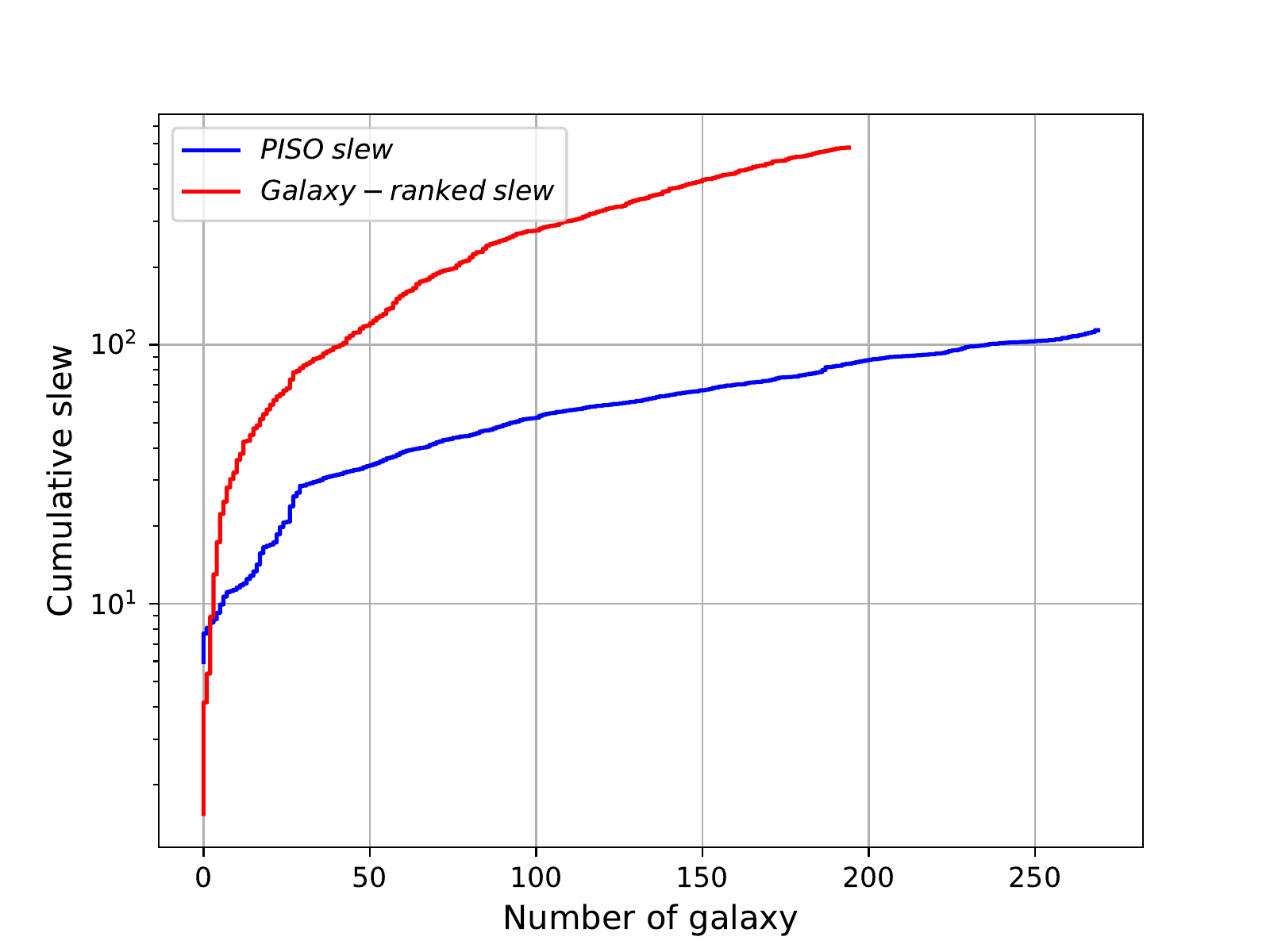}
 \caption{Comparing \iso\ and \grm\ method. The plot shows the comparison of the probability coverage of the two methods. The X--axis is the number of galaxies in all pointings, and the Y--axis shows the cumulative probability covered observing all those galaxies. The \iso\ method is shown in blue and the \grm\ method is shown in red color.
 Compare the total slew angle for \iso\ and \grm\ method. The plot shows the comparison of the cumulative slew coverage of the two methods. The X--axis is the number of pointings, and the Y--axis shows the cumulative slew angle for observing all those pointings. The \iso\ method is shown in blue and the \grm\ method is shown in red color.
     \label{fig:compare_probability}}
\end{figure*}
\makeatletter\onecolumngrid@pop\makeatother

\begin{figure}[thbp]
  \centering
    \includegraphics[trim=0cm 0 0cm 0cm,clip=false,width=0.3\textwidth]{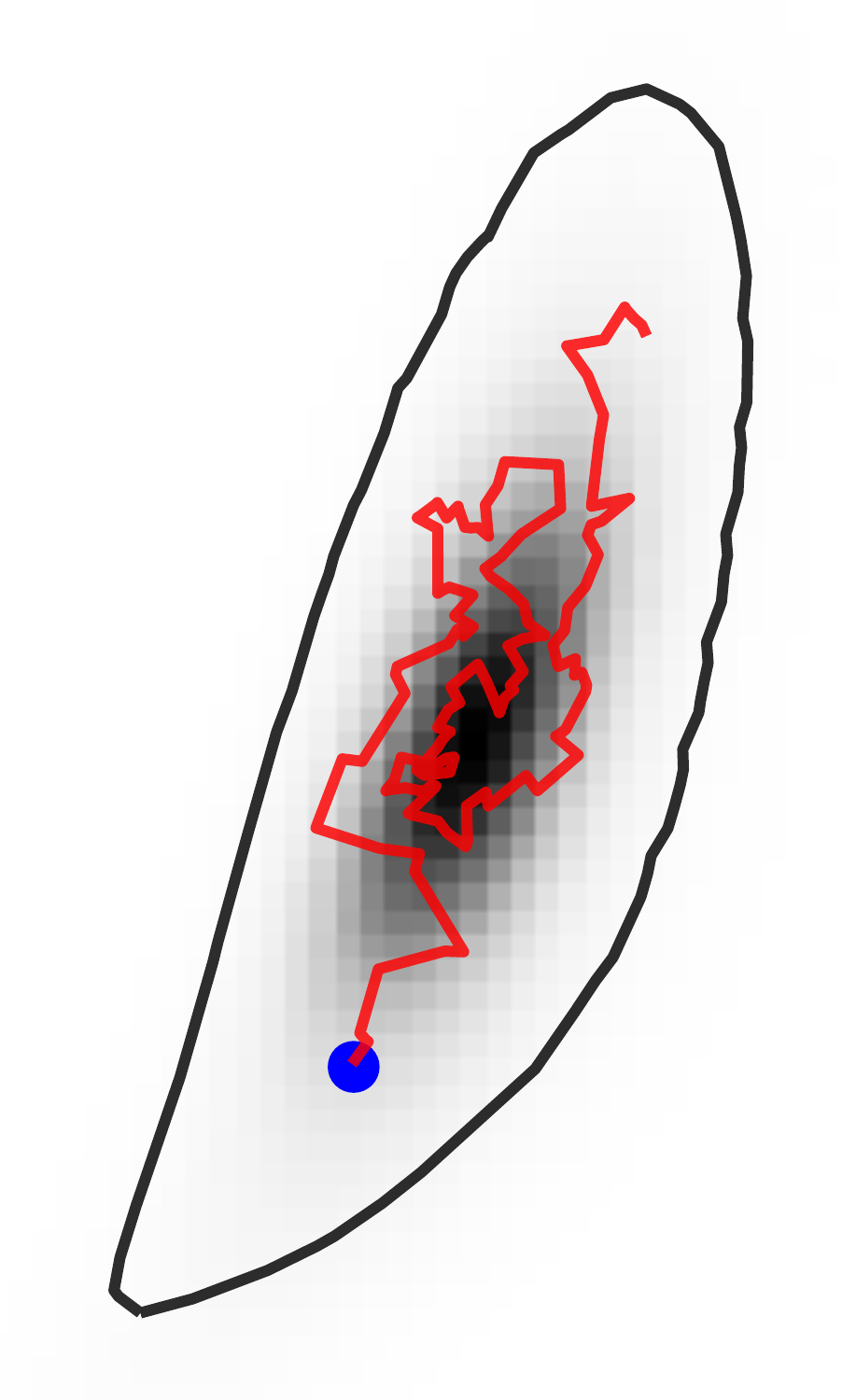}\hspace*{0.5cm}
  \caption{Sky view of the slew path for the observation of the 112 sq.deg. localization (at a mean distance of 106 Mpc) described in \S\ref{sec:application:single} and Figure~\ref{fig:compare_probability}. The grey shade is the GW-probability distribution and the black curve is the contour of the 90\% probability containment area on the sky. The blue dot is the initial position of the radio antennas and the red lines are the optimized slew paths of the radio antennas by \iso\ method. For details, see \S\ref{sec:application:single}.}
    \label{fig:sky_slew}
\end{figure}

In the following description of the comparisons, we assume VLA antennas observing at S band\footnote{We consider the VLA 3 GHz primary beam up to the 10\% power point ($11^{\prime}$ radius). The beam sensitivity pattern is from \cite{Perley2016bqt}, i.e. $b(\alpha) = 1-1.505\times 10^{-3} \times \alpha^2 + 8.37 \times 10^{-7} \times \alpha^4 - 1.75 \times 10^{-10} \times \alpha^6$ in equation~\ref{equ:beam_int} (2--4 GHz). We assume the integration time to reach a rms noise of 20 $\mu$Jy is 240 seconds. The antenna slew rate is taken to be 40 deg/min in azimuth and 20 deg/min in elevation.} seconds per degree. 
We assume a limiting luminosity of the radio afterglow search such that a source at 200 Mpc will be detected with a 4$\sigma$ confidence with 240 sec integration time.
We have assumed the total ToO time for one epoch observation is 7.5 hours. 

\subsection{Comparison between \iso\, and the simple \grm\  methods}\label{subsec:comparison}
In all our comparisons we treat slew time independently and assume that the total time integration time available for observing galaxies is 7.5 hr.

\subsubsection{A single 112 sq.deg GW localization at 106 Mpc} \label{sec:application:single}
Here, we compare between the results from the \iso\, and the \grm\ methods for a single single 112 sq.deg localization at a median distance of  106 Mpc. The volume of the 90\% probability containment region is 2.3$\times 10^4$ Mpc$^3$ and the total number of known galaxies within the volume is 943.

We first look at the improvement from the pointing and integration time optimization of the \iso\ method.
In the left panel of Figure~\ref{fig:compare_probability} we plot the cumulative probability covered as a function of the number of galaxies observed. We see that the \grm\ method covers about 300 galaxies, reaching a cumulative containment probability of 53\%. Although the containment probability covered by the \iso\ method with 300 galaxies is lower compared with the \grm\ method, the \iso\ method is able to cover $>$500 galaxies (in the given observing time), reaching a cumulative containment probability of 60\%.
In terms of containment probability the \iso\ method (blue curve) therefore outperforms the \grm\ (red curve) method.  

For the 112 sq.deg. localization we now calculate the total slew for the two methods. In the right panel of Figure~\ref{fig:compare_probability}, \grm\ method schedules the the pointings such that the total amount of slew is $\sim$578 degree (red curve), where \iso\ method optimize the slew and improve tremendously giving $\sim$114 degree (blue curve) of total slew. The \iso\ method thus saves $\sim$15 min of overhead (slew time) compared to the simple \grm\ method. 

In Figure~\ref{fig:sky_slew}, we show the path of the antennas for all the pointings selected by the \iso\ method. The grey shaded region is the GW-probability distribution and the black contour defines the area of the 90\% probability containment on the sky. The blue dot is the initial position of the pointing of the radio antennas (first pointing, chosen due to its earliest setting time) 
and the red lines are the optimized slew paths of the radio antennas by \iso\ method. 

\subsubsection{Comparisons for the $\sim$1200 GW localizations accessible to the VLA} \label{sec:application:all}
\begin{figure*}[t]
\centering
\includegraphics[trim=0cm 0 0cm 2cm,clip=true,width=\textwidth]{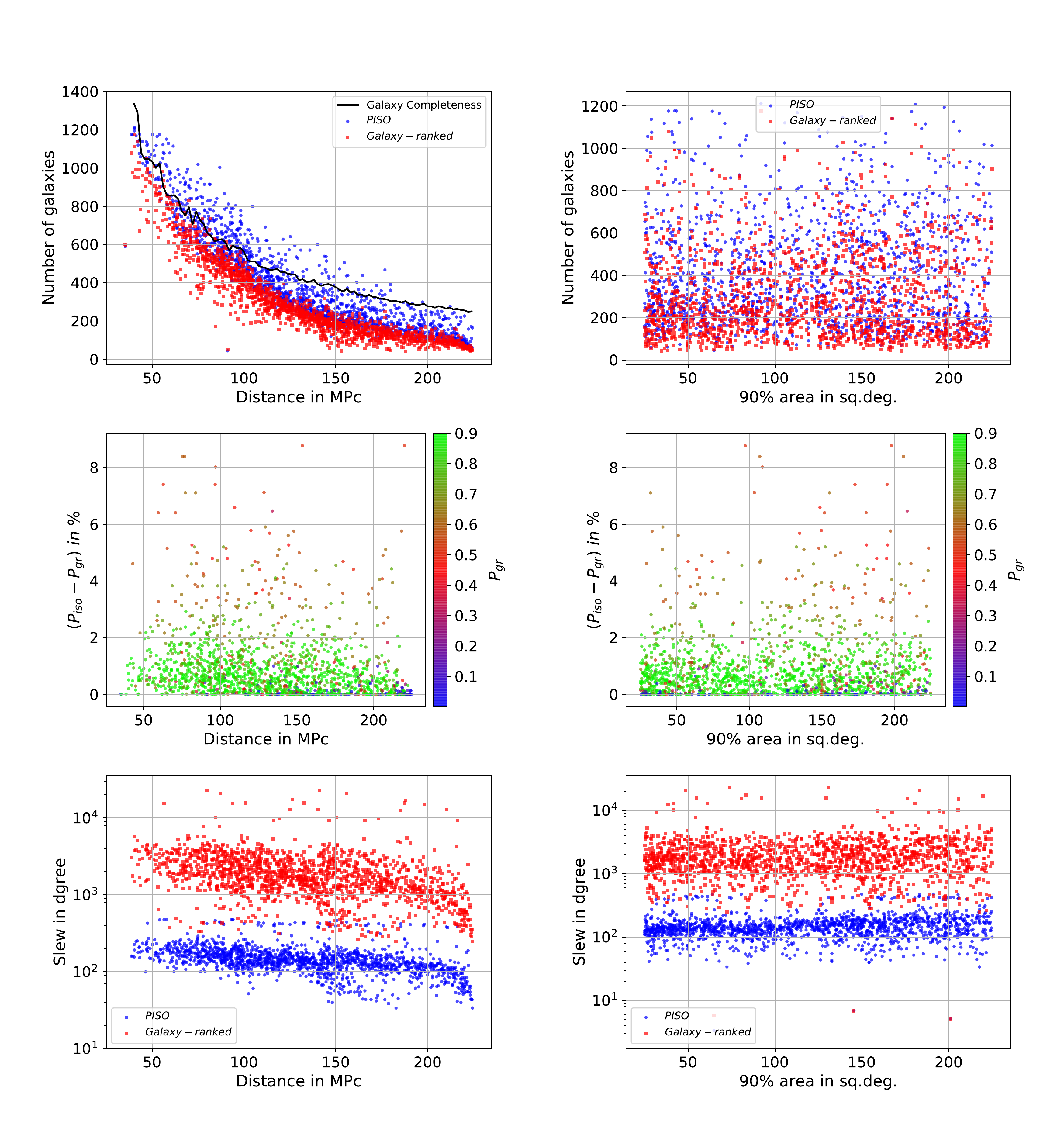}
 \caption{Comparisons between the \iso\ and \grm\ methods for $\sim$1200 simulated GW localizations that are observable with the VLA. The \iso\ method performs better than or equal to the \grm\ method for all the cases, as expected. {\it Top panels}: the results from the \iso\ and \grm\ methods are represented by blue and red points respectively. The X--axes in the left and right panels are the mean distance (from Earth) and the 90\% probability containment area of the localizations respectively. The Y--axes shows the total number of galaxies covered within the localizations by the two methods. The black curve in the left panel shows the number of galaxies covered by \iso\ method at different distances assuming a roughly complete catalog (see \S\ref{sec:application:all} for details). \textit{Middle panels}: These panels give the relative probability improvement, for the \iso\ method relative to the \grm\ method, as function of distance (left-panel) and sky localization area (right-panel). The color coding represents the cumulative probability within the GW localization map covered by the \grm\ method. Note that the maximum covered probability is 0.9, as we consider the patches in the simulation are the 90\% probability containment area of the localizations. \textit{Bottom panels}: The comparison of the total slew between the \iso\ and \grm\ methods for $\sim$1200 localizations, accessible to the VLA, is shown. Color coding is the same as that in the top panels.}
  
\label{fig:sim-plots}
\end{figure*}

In Figure~\ref{fig:sim-plots}, we show the comparisons between the \iso\ and \grm\ methods for all the $\sim$1200 GW localizations observable with the VLA. The \iso\ method performs better than \grm\ method for most of the patches. In none of the cases do we find that the \iso\ under-performed compared to the \grm\ method. Figure~\ref{fig:sim-plots} shows the improvement \iso\ method in covering the number of galaxy and the total probability within those localizations compared to the \grm\ method. Each event observed by \iso\ and \grm\ are represented by a blue and red points respectively in the top and bottom panel. The X--axes in the left and right panels are the mean distance and the 90\% probability containment area of the localizations respectively. 

In the top panels, Y--axes shows the total number of galaxies covered within the localizations by the two methods. 
In the top-left panel, the number of galaxies covered per event decreases with distance as the number of pointings within a given ToO time reduces with distance. The number of pointings per event will be lesser at higher distance, because the integration time increases as fourth power of the distance. 
The \iso\ method covers between 10--200 galaxies more than the \grm\ method, the mean being $\sim$95. 

To understand the effect of galaxy catalog incompleteness, we used the following crude approach. 
We assumed the catalog to be 100\% complete within 50 Mpc (distance from Earth), and 
calculated the average number density ($n_{{\rm gal},50}$ Mpc$^{-3}$) of galaxies within this distance.
We then prepared a simulated galaxy catalog enforcing the number density of galaxies between 40--300 Mpc to be the same as $n_{{\rm gal},50}$. We used this simulated galaxy catalog to test the performance of \iso\ method on several GW localizations. The black curve in the top-left panel of Figure~\ref{fig:compare_probability} represents the number of galaxies covered by \iso\ method at different distances between 40--225 Mpc with the simulated galaxy catalog. 
In this case, we have used the GLADE \citep{dalya2018}, and the effect of galaxy catalog incompleteness beyond 100 Mpc is evident in the top-left panel. 

\makeatletter\onecolumngrid@push\makeatother
\afterpage{\clearpage}
\clearpage
\makeatletter\onecolumngrid@pop\makeatother

The middle panels of Figure~\ref{fig:sim-plots} show the relative probability improvement as function of distance (left-panel) and area (right-panel). In the middle panel, the Y--axes are representing the additional probability gained in percentage by the \iso\ method than the \grm\ method. The color coding represents the cumulative probability within the GW localization map covered by the \grm\ method. The maximum covered probability is 0.9, as we consider the patches in the simulation are the 90\% probability containment area of the localizations. 
Containment probability gained by the \iso\ method ranges between 0--9\%. We note here again that this gain in containment probability is only due to the pointing and integration time optimization within the \iso\ method. We have not considered slew time to count towards the total allocated observing time $T_{too}$. The optimization of slew time (results described in the following paragraph) will substantially enhance the gain in probability.

The bottom panels of Figure~\ref{fig:sim-plots} shows the comparison of the total slew between the \iso\ and \grm\ methods for the $\sim$1200 observed localizations. The X--axes of the bottom--left and the bottom--right panels are the mean distance and the 90\% probability containment area of the localizations respectively. The Y--axes in both the bottom panels are the total amount slew of the radio antennas in degrees. The optimization of the slew by \iso\ method provides an order of magnitude improvement than the slew of \grm\ method. The declining pattern of the slew with increasing distance in the bottom-left panel is because the total number of pointings per localization reduces with increasing distance, as alluded to earlier. The average improvement in the total slew by \iso\ method than \grm\ method for the $\sim$1200 localizations is $\sim$1800 degrees. This corresponds to $\sim$55 minutes (i.e. 12\% of the total 7.5 hours considered for each epoch) of slew time for the VLA. 

\begin{table*}
\caption{Comparison between the \iso\ and \grm\ methods for GW localizations at particular distances.}
\begin{tabular}{|c|c|c|c|c|c|c|c|c|c|}\hline
  \multirow{2}{*}{90\% area} & \multirow{2}{*}{Mean distance} & \multirow{2}{*}{Sigma distance} & \multicolumn{3}{|c|}{\textbf{\iso}} & \multicolumn{3}{|c|}{\textbf{\Grm}}  \\ \cline{4-9}
 & & & Probability & No. of Galaxies & Slew$^*$ & Probability & No. of Galaxies & Slew$^*$ \\
 in sq.deg. & in Mpc & in Mpc & covered$^*$ &  & in deg. & covered$^*$ &  & in deg. \\ \hline

48  & 40  & 13 & 90.0 & 805 & 358 & 90.0 & 758 & 4237 \\ \hline
72  & 80  & 21 & 78.1 & 613 & 284 & 73.4 & 521 & 3748 \\ \hline
123 & 120 & 52 & 69.3 & 393 & 153 & 61.3 & 260 & 906  \\ \hline
139 & 160 & 70 & 63.8 & 225 & 104 & 55.6 & 127 & 479  \\ \hline
187 & 200 & 73 & 49.5 & 178 & 53  & 42.5 & 118 & 265  \\ \hline
\multicolumn{8}{l}{$^*$Assumes 7.5 hr total integration time. Slew time is treated independently.}
\end{tabular}
\label{table:dist-test}
\end{table*}

\subsubsection{Comparisons for GW localizations at different distances} \label{sec:application:distance}

In Table~\ref{table:dist-test}, we present the comparison of the two methods, \iso\ and \grm, for five representative localizations at different distances. 
For the localizations at smaller distances (having area $\lesssim$100 sq. degrees), the total probability coverage and the total number of galaxies observed are somewhat similar for the two methods, but the slew is substantially different (by an order of magnitude).
At larger distances (and correspondingly larger localization areas), the containment probability covered by the \iso\ method is significantly better than the \grm\ method, while the improvement in terms of slew overhead is somewhat smaller than in the small-distance case.  
The number of galaxies observed in the two methods is comparable, since 1) the increase of the integration time per pointings at higher distance, which reduces the total number of pointings per event for the fixed amount of ToO time, and 2) the incompleteness of the galaxy catalog at higher distance. 
We note that the convolved probability is normalized over the known galaxies in the catalog.

Since galaxy catalog incompleteness can be a significant issue at distances $\gtrsim$150 Mpc, we explore a means to mitigate the effect in the following section.

\section{Mitigating the effect of galaxy catalog incompleteness}\label{sec:mitigation}
The completeness of the galaxy catalog decreases with distance within the range of LIGO sensitivity for BNS mergers. The NASA Extragalactic Database (NED) catalog is around 78\% complete at 130 Mpc \citep{kulkarni2018}. As the pointing order in \iso\ and \grm\ methods depend on the galaxies within the localization, the pointings might be biased as the galaxy catalogs are not complete. We show below that the bias could be mitigated (especially if the final probability map is weighted by the galaxy mass) through pointing towards the higher mass galaxies and/or including fields flanking the higher mass galaxies. These observations may cover a significant number galaxies that are missing in the catalog.

\begin{figure}[thbp]
  \centering
    \includegraphics[trim=2cm 0 0cm 0cm,clip=false,width=0.55\textwidth]{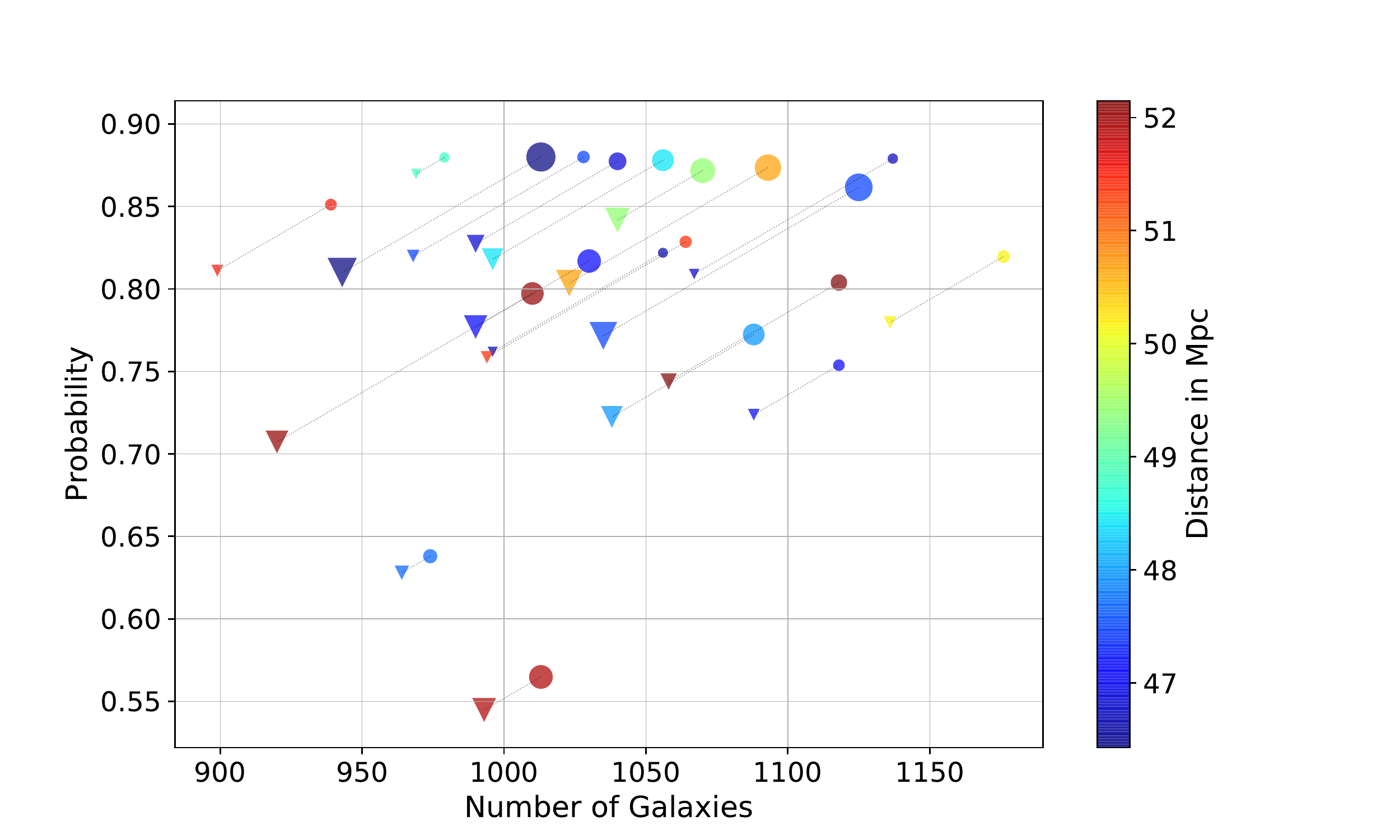}\hspace*{0.5cm}
  \caption{Results of a test to find the effect of galaxy catalog incompleteness on the counterpart search. The X--axis represents the number of galaxies covered within a localization and he Y--axis represents the probability covered by those galaxies. \textit{Circular Points:} The case when all the galaxies within the localization from the galaxy catalog are used to test the \iso\ algorithm. \textit{Triangular Points:} The case when half of the galaxies within the localization are removed based on the luminosity of the galaxies. The size of the scattered points are proportional to the 90\% area of the localizations. The color code shows the mean distance of the localizations. See \S\ref{sec:mitigation} for details.
    \label{fig:pnda}}
\end{figure}

In order to find the bias and attempt to observe the galaxies that may be missing in the galaxy catalog, we begin with the assumption that the catalog is $\sim$100\% complete (in terms of mass and star formation rate; for this exercise we use the GLADE catalog) at a distance of $\sim$50 Mpc. We run a test for a set of 30 localizations to check the bias of the \iso\ algorithm for two scenarios. In one scenario, the pointings of the observation are done keeping all the galaxies within the localization intact. The other scenario, half of the galaxies within the localization are removed based on their lower B-band luminosity, while scheduling the pointings. We choose the luminosity threshold to remove half of galaxies within the localization. The threshold luminosity might vary for different localization. On an average $\sim$500 galaxies were removed per localization. But, when we calculate the beam probability and the number of galaxies within the antenna beam, we include the removed galaxies if they are inside the beam. This implies that, if there are any galaxies that are absent in the catalog but falling inside the beam when the antennas are pointing to a cataloged galaxy, then the observation of those unknown galaxies will reduce the bias. In Figure~\ref{fig:pnda}, we show the result of this test for a subset (20) of the GW localization regions searched. The X--axis represents the number of galaxies covered within a localization and the Y--axis represents the probability covered by those galaxies. Circular points indicate the cases where all the galaxies within the localization from the galaxy catalog are used to schedule the observation by \iso\ algorithm and triangular points show the cases where the pointings are optimized after half of the galaxies within the localization region are removed. The size of the scattered points are proportional to the 90\% area of the localizations. 
The dotted lines join a triangular point and the corresponding circular point. Through this exercise we find that (as expected) many of the low-mass galaxies are located close to higher mass galaxies, and end up being observed in pointings towards those high mass galaxies. 
We also find that the difference in the probability covered between the two cases (i.e. the displacement in the circular and corresponding triangular points in the Figure) is significant, but less than 10\%.
This implies that observing pointings near high mass galaxies in the localization region is expected to cover a significant fraction of the uncataloged galaxies and thereby reduce the pointing bias of the \iso\ method.
As the GW source distance (from Earth) increases, the number of galaxies within each antenna beam will also increase, reducing the pointing bias even further.
Here we have assumed VLA antennas, having beam sizes of a few arcminutes.
For larger beams (e.g. MeerKAT), the bias due to galaxy catalog incompleteness is expected to be much lower.

\section{Summary \& Discussion}\label{sec:summary}
We have presented an algorithm to optimally search for the radio afterglows of GW events using the galaxy targeting approach.
The algorithm, described in \S\ref{sec:algo} (mathematically in \S\ref{sec:algo:math}--\S\ref{sec:algo:slew}; as a flowchart in \S\ref{sec:algo:flowchart} and Figure~\ref{fig:flowchart}), involves optimizing the a) location of antenna pointings, b) the integration time, and the c) total antenna slew, in order to maximize the observed galaxies (weighted according to mass or other parameters) and the containment probability.
We refer to this approach as the pointing, integration-time and slew optimization (\iso\ method). 
This method performs a search down to a particular (user-specified) limiting luminosity for each galaxy observed in the 90\% GW containment region; this approach is useful for placing meaningful constraints on the physical, microphysical, and ISM parameters related with the afterglow.

We simulated $\sim$2000 GW localizations (\S\ref{sec:sim}) in order to test the performance of our algorithm.
Comparing our algorithm to the simple method of sequentially targeting a rank-/weight-ordered galaxy list (\grm\ method) we find that the pointing and integration time optimization of \iso\ alone can gain a few percent in terms of containment probability, and further cut down the slew overheads substantially with through antenna slew optimization. 
Taken together, the \iso\ method allows $\sim$10\% gain in containment probability compared to the \grm\ method.
The application of the algorithm to a single 112 sq. deg localization at 106 Mpc (\S\ref{sec:application:single}), to all simulated events (\S\ref{sec:application:all}), and to a few events at different distances (\S\ref{sec:application:distance}) assuming a VLA search at S band. 
Further, we find that the improvement in slew overheads is enormous ($\sim$1800 degrees on an average for all the simulated localizations), comparing the \iso\ method to the \grm\ method. 

During LIGO/Virgo run O3 and beyond, the mean distance at which NS-NS/NS-BH mergers ($\gtrsim$150 Mpc) are detected will increase to a point where galaxy catalog incompleteness becomes significant.
The galaxy-targeting approach may therefore miss some possible host galaxies.
In order to observe those galaxies that may be missing in the galaxy catalogs (and to increase the completeness of the search, especially in terms of mass), we suggest doing deeper observations towards the higher mass galaxies in the 3D GW localization region, and/or including fields flanking these higher mass galaxies.
In \S\ref{sec:mitigation} we calculate the difference in the number of galaxies observed, and in the containment probability covered by the radio search, when dealing with an incomplete and complete catalogs. 
The results are shown in Figure~\ref{fig:pnda}, where we find the difference in the probability covered between the two cases is significant, but less than 10\%.

Over the next few years, the detection of EM counterparts of mergers will enable $\sim$arcsec localization and host galaxy identification.
This will be important for determining the weighting of galaxies within the GW localization regions, which is expected to play an important role in galaxy-targeted searches of the EM counterparts.
Further improvements in our search method are possible by choosing an optimal galaxy weighting scheme.

\quad\newline
\noindent Acknowledgements: We thank Shri Kulkarni and Eran Ofek for suggestions that helped improve this paper. KPM is currently a Jansky Fellow of the National Radio Astronomy Observatory. 

\software{Matplotlib~\citep{hunter2007},~NumPy~\citep{numpy}, 
Astropy \citep{astropy2018}, }

\bibliography{mybib.bib}

\end{document}